\documentclass[acmlarge,nonacm]{acmart}
\usepackage{adjustbox}
\usepackage{phdcode}

\begin{document}

\title{Managing Research the Wiki Way}
\subtitle{A Systematic Approach to Documenting Research}

\author{José Devezas}
\email{jld@fe.up.pt}
\orcid{0000-0003-2780-2719}

\author{Sérgio Nunes}
\email{ssn@fe.up.pt}
\orcid{0000-0002-2693-988X}

\affiliation{%
  \institution{INESC TEC \& Faculty of Engineering, University of Porto}
  \streetaddress{Rua Dr. Roberto Frias, s/n}
  \postcode{4200-465}
  \city{Porto}
  \country{Portugal}
}

\begin{abstract}
  As a master's student, knowing how to manage your personal research is not only useful for keeping track of your work, but it is also a process that should be learned as a part of your training. As a doctoral student, however, research management is a fundamental part of your overall methodology and it should be a well-planned process. Long-term research requires a good approach to documentation, otherwise you risk getting lost among your many surveyed papers, carried experiments, and results. This approach should be systematic, accessible (mainly to you), low-effort, and a natural part of your daily workflow --- it should be there to help you, and not the other way around. In this article, we describe how we relied on a wiki to organize literature, datasets, experiments and results, and we also show how such a systematic approach can lead to better insights through automated meta-analysis. In addition to this content, we provide a docker-based installation of a preconfigured wiki, with the required templates and extensions, along with some examples pages, as well as a Jupyter notebook to analyze your documented work. So read on.
\end{abstract}


\maketitle

\section{Introduction}
\label{introduction}

As a computer scientist, and particularly when focusing on long-term research, it is fundamental to devise a strategy for the personal documentation of the process. Our team has, more than once, relied on a simple DokuWiki\footnote{\url{https://www.dokuwiki.org/}} solution for this purpose. DokuWiki is a PHP file-based wiki that is easy to deploy and backup (just compress and archive it, or synchronize a copy to you local machine), while retaining historical changes to pages and having access to easy extensibility. In order to organize our work, we created templates to help document everyday tasks, such as literature review, or experiment planning and result archiving. These templates supported the creation of reading or experiment sheets, which heavily promoted the use of links. This enabled us to create special pages containing automatic indexes by author, conference or journal, or to simply find which experiments a collection was used in by visiting the collection's page. Moreover, the wiki is searchable, and it provides redundancy against content that you might want to recover, as you are writing and planning. We have previously introduced this work in Devezas and Nunes~\cite{Devezas2021}. In this article, we provide an extended description of our systematic documentation approach, carried in the context of the first author's doctoral work in graph-based entity-oriented search, and running an automatic characterization of the literature, as annotated in the wiki.

\section{Systematic documentation}
\label{systematic-documentation}

We relied on DokuWiki for documenting doctoral work over the course of four years. We systematically organized information about literature, collections, and experiments, providing templates for reading sheets, collection descriptions, and experiment note-taking and result archival. We thoroughly exploited links and backlinks, establishing relations between authors or conferences and their publications, or datasets (collections) and their mentions in other pages. We also linked subsets to their original dataset, or related experiments among themselves. Our systematic documentation methodology consists of three parts, which we describe next. First, we describe the literature review methodology along with the reading sheet creation process. Next, we cover a lightweight approach for documenting collections based on a description sheet template, which includes information about subsets and evaluation results taken from the literature. Finally, we describe a note-taking and archival strategy for experiments and their results.

\subsection{Literature review: linking reading sheets}
\label{literature-review-linking-reading-sheets}

We relied on an exploratory literature review approach, focusing and refining along the process, as concepts became clearer. We used academic search engines to issue queries in an attempt to solve our information needs about entity-oriented search or network science approaches that could be useful for graph-based models. Resulting publications were selected by reading the title, the abstract, the conclusions, and sometimes a part of the introduction, in this order. Selected publications were then added to the wiki, along with a list of specific goals in the form of tasks, to be reviewed in order of priority regarding ongoing research work, or based on the overall relevance and informational value to the thesis. A reading sheet was created for each publication, containing a standardized table of information, as well as reading notes organized according to the structure of the sections of the publication. We frequently included block quotes highlighting important information, and we always finished with a summary paragraph of the work. Next, we describe this workflow in further detail, step by step.

\begin{figure}
  \centering
  \includegraphics[width=\linewidth]{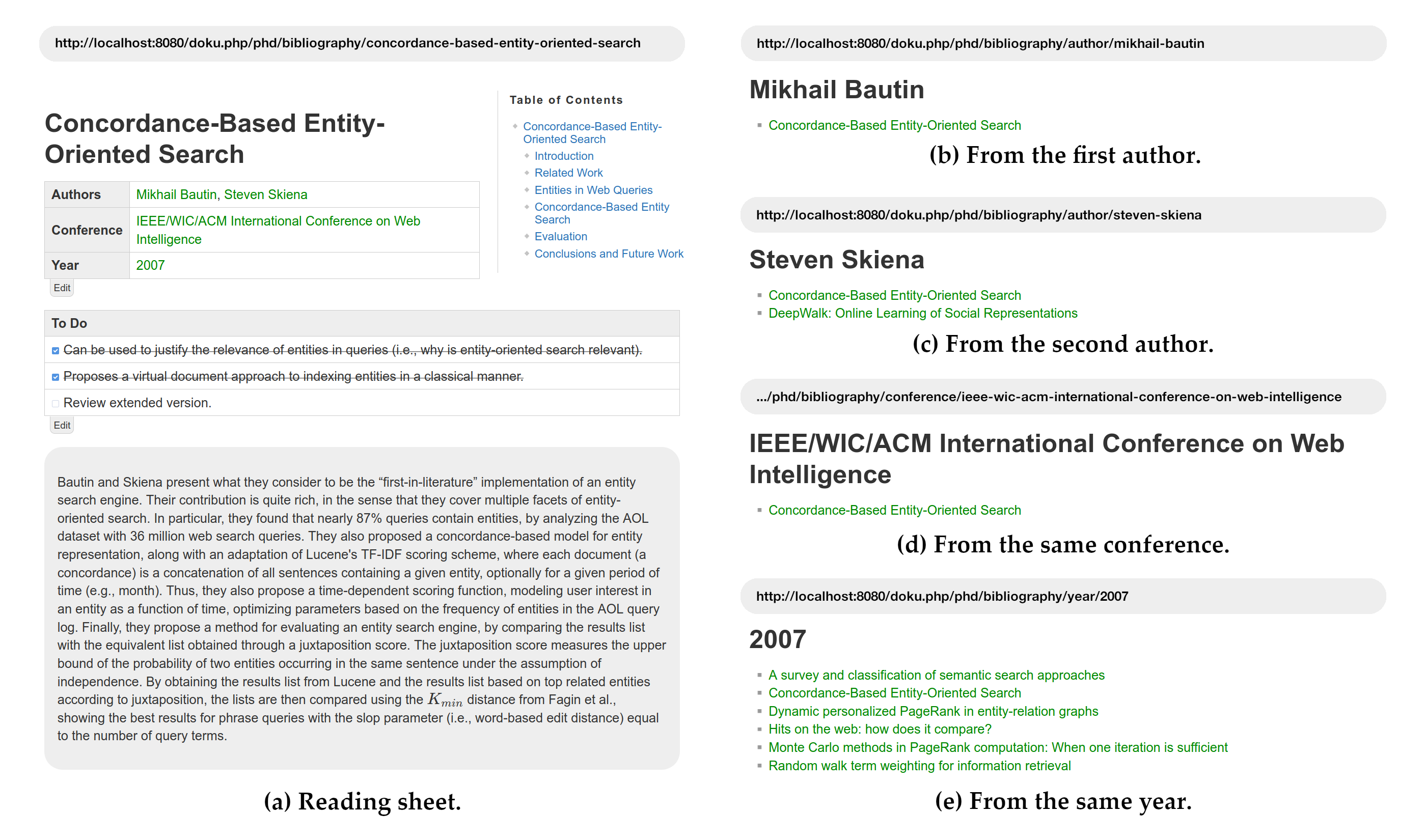}
  \caption{Systematic documentation of reviewed literature.}
  \label{fig1}
\end{figure}

\begin{enumerate}
  \item \textbf{Search Google Scholar}\footnote{\url{https://scholar.google.pt/}} and/or \textbf{Semantic Scholar}\footnote{\url{https://www.semanticscholar.org/}} based on a set of queries about your research topic.

  \item \textbf{Follow Google Scholar alerts} for high granularity queries, tailored to match relevant state-of-the-art contributions. And follow \textbf{Semantic Scholar alerts} for the topics of interest to you.

  \item \textbf{Do a quality assessment}, selecting publications based not only on the title, abstract and conclusions, but also on whether they are indexed by well-known bibliographic databases (e.g., \href{http://dblp.org/}{DBLP}, \href{http://dl.acm.org/}{ACM Digital Library}).

  \item \textbf{Review priority publications} on the broad subjects of the thesis, including relevant surveys, and conference or journal articles, that cover relevant or similar approaches (i.e., graph-based approaches to search, or the combination of corpora and knowledge bases).

  \begin{enumerate}
    \item Each reviewed publication is added to the wiki, and a reading sheet is created in a page named with the lowercase, dash-separated title of the publication, discarding non-alphanumeric characters (e.g., \textit{phd:bibliography:concordance-based-entity-oriented-search}).

    \item Each reading sheet contains a table with links to author pages, a bibliographic collection page (i.e., journal, conference, institution) and a year page, following the same naming convention as described above.

    \item Each author, collection and year page contain a list of all backlinks to that page (i.e., all the entries by an author, in a collection, or published in the given year). Optionally, these pages can also include relevant information about the author, the collection or even the year (e.g., a summary of the discoveries or state of the art in that year). We follow the naming convention from item 1 for the author pages, using the full name as it appears in the reviewed publication, including single letters (e.g., \textit{phd:bibliography:author:w-bruce-croft}).

    \item Below the reading sheet table, we replicate the publication's sections and take notes on each section during reading.

    \item After reading a publication, we insert a written summary of the whole publication below the reading sheet table (and above the notes).

    \item Links to reviewed publications are displayed in bold, while links to ongoing or short-term reviews are prefixed with a label for either \textbf{{[}In Review{]}} or \textbf{{[}To Review{]}}, which respectively link to \textit{phd:bibliography:in-review} and \textit{phd:bibliography:to-review}, maintaining lists of publications being worked on and scheduled for future review.
  \end{enumerate}

  Figure~\ref{fig1} illustrates the type of content created during the review process, showing the reading sheet (Figure~\ref{fig1}a), the list of all articles from the first and second authors, available in the wiki (Figures~\ref{fig1}b and \ref{fig1}c), and the list of articles from the same conference and year (Figures~\ref{fig1}d and \ref{fig1}e). We also display the resulting URL for each page.

  \item \textbf{Group selected publications} into relevant subjects and briefly summarize all publications, delving into more detail as deemed relevant.

  \begin{enumerate}
    \item Compile written summaries into a logical sequence.
    \item Contextualize each publication in regard to the thesis.
  \end{enumerate}
\end{enumerate}

\subsection{Collections: metadata, samples and evaluations}
\label{collections-metadata-samples-and-evaluations}

\begin{figure}
  \centering
  \includegraphics[width=.75\linewidth]{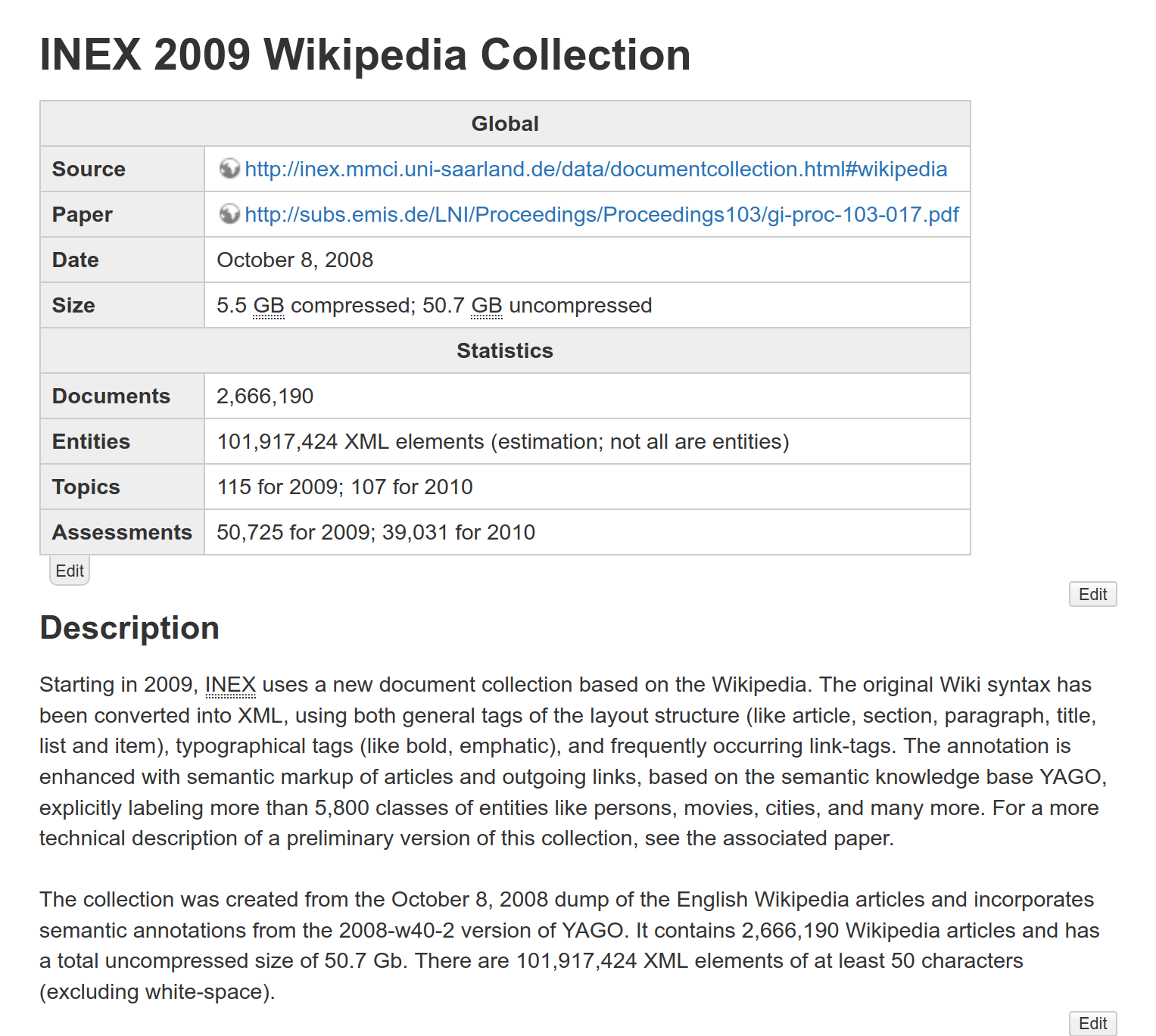}
  \caption{Systematic documentation of the INEX 2009 Wikipedia
  collection.}
  \label{fig2}
\end{figure}

We created a page in the collections section of the wiki for each dataset that we used, created, or otherwise explored. In order to ensure consistency in the description, we prepared a template containing a table of metadata to be filled about each collection, along with a longer textual description. The fields that we considered were the following: `Source', `Paper', `Date', and `Size'. We also added fields for statistics specific to test collections within the domain of entity-oriented search: `Documents', `Entities', `Topics', and `Assessments'. For other types of data, these statistics were ignored or replaced. For example, for network data, we used `Nodes' and `Edges' instead. Figure~\ref{fig2} illustrates the template that we have just described, in its application to the INEX 2009 Wikipedia collection.

Due to scalability issues, we also relied on multiple samples of our main test collection. Each sample was described in the wiki as any other collection, assigning a link, to the corresponding descriptor, pointing to the page of the original dataset, instead of what would otherwise be a URL to the source website or similar resource.

Whenever available, we also included evaluation results, found within the literature, when applied to a task of interest to us, based on the same dataset. This information is usually available in overview papers from the evaluation forums that provide the test collection.

\subsection{Experiments: taking notes and archiving results}
\label{experiments-taking-notes-and-archiving-results}

\begin{figure}
  \centering
  \includegraphics[width=\linewidth]{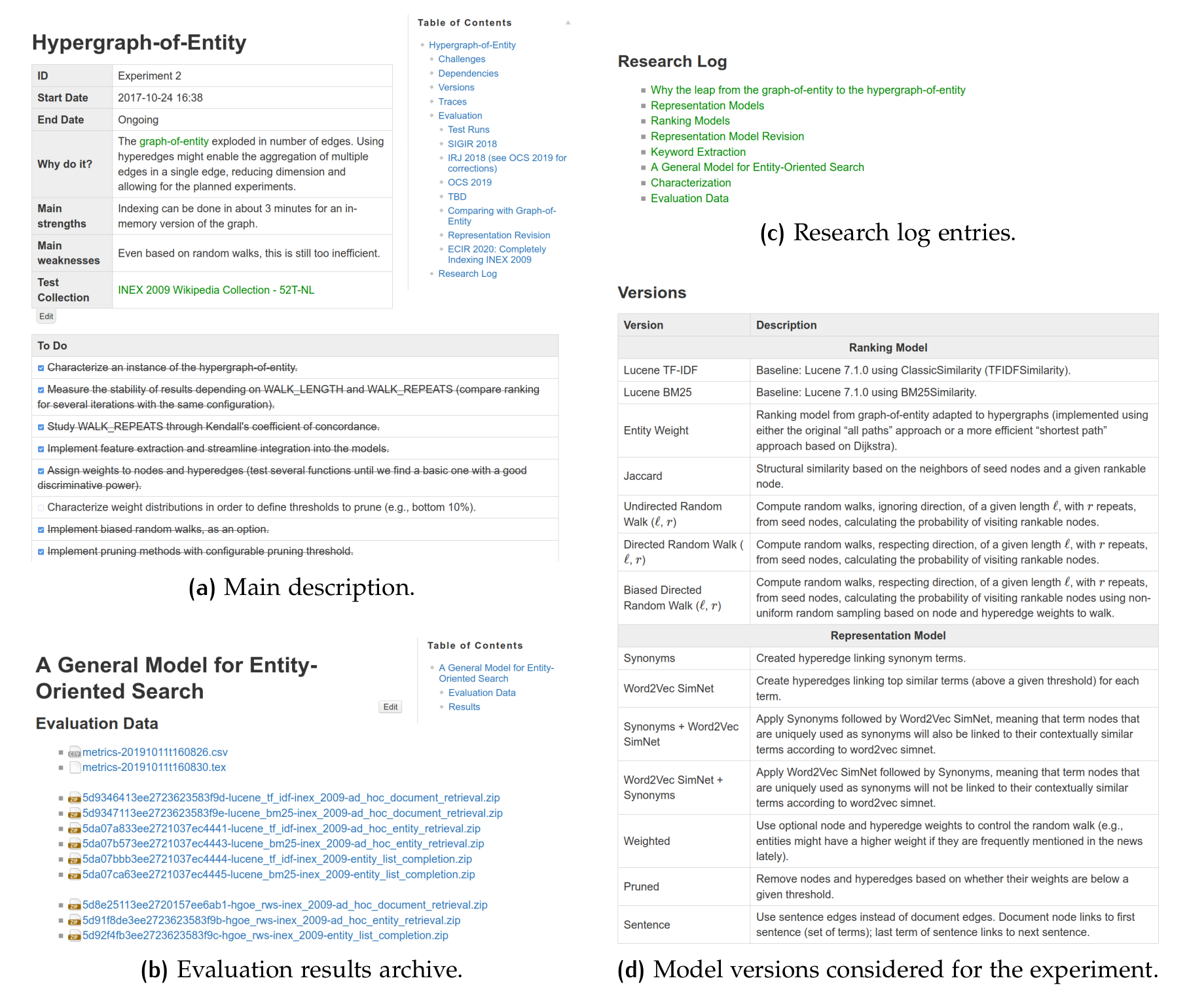}
  \caption{Systematic documentation of the hypergraph-of-entity
  experiments.}
  \label{fig3}
\end{figure}

Following a similar philosophy to the literature review and collection description, we also established a template for the documentation of experiments in the wiki. Each experiment page contains a metadata table (Figure~\ref{fig3}a) with the following fields: `ID' (e.g., ``Experiment 1''), `Start Date' (e.g., ``2017-10-24 16:38''), `End Date' (e.g. ``Ongoing''), `Why do it?' (motivation; usually in the sequence of a previous experiment), `Main strengths' (expected improvements), `Main weaknesses' (expected issues), and `Test collection' (a link to a wiki collection page).

Each experiment also included a to-do list, a description table of the model versions explored (Figure~3d), and an evaluation section with performance metrics for each of those versions. Additionally, we also added tables to describe predicted or identified challenges, and dependencies that would block the execution of the experiment. Over time, the challenges and dependencies sections were mostly deprecated in favor of the to-do list, and the archival of trace logs proved cumbersome and with little utility, thus being increasingly ignored over time. On the other side, the archival of results from evaluation tasks (Figure~\ref{fig3}b) was a particularly useful practice that enabled us, for instance, to verifying the calculation of effectiveness metrics.

Finally, we created a section for research logs (Figure~\ref{fig3}c), where we added links to wiki pages under the hierarchy of the current experiment. Each research log entry represented a reflection on parts of studied models, possibly branching into follow-up or sub-experiments.

\section{The wiki way}
\label{the-wiki-way}

Wikis have been used effectively in learning environments~\cite{Ruth2009}. Following the previously described methodology, we prepared a DokuWiki instance with a series of templates, spread over a predefined namespace structure that would support and facilitate research management in a flexible way. We also installed several useful extensions, a list that we compile here as the result of several years of active curation. Finally, we provide a docker-based installation, as well as the option to simply rely on the wiki skeleton that we prepared to integrate in your existing DokuWiki instance.

\subsection{Organizational structure}
\label{organizational-structure}

\begin{figure}
  \centering
  \includegraphics[width=\linewidth]{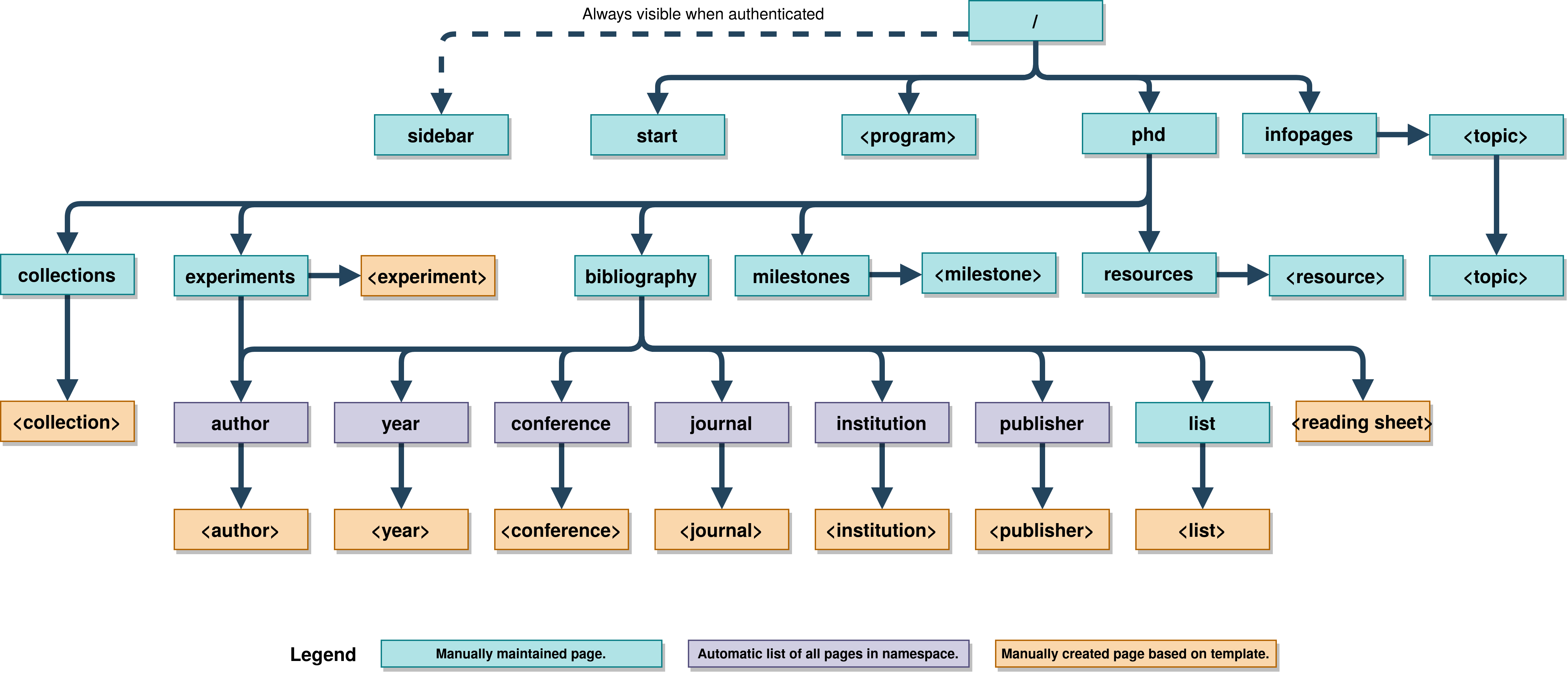}
  \caption{Page structure for the research management wiki.}
  \label{fig4}
\end{figure}

We established a wiki structure purely based on need. The organization we share next is the result of four years of reiterating and reorganizing based on these personal requirements. Pages are listed with the extension \textit{.txt} (they are text files in the filesystem), while namespaces have no extension (they are directories in the filesystem). The overall page structure is also summarized in Figure~\ref{fig4}.

\begin{itemize}
  \item
    \textbf{sidebar.txt:} Default DokuWiki page that contains the menu on
    the left, only visible while logged in.
  \item
    \textbf{start.txt:} Default DokuWiki landing page. This is the only
    public page by default. It can be used to publicly communicate
    information about yourself and your research.
  \item
    \textbf{infopages.txt:} Then entry page, presenting an organized view
    over explored topics.
  \item
    \textbf{infopages:} Your own personal ``Wikipedia'', but instead of
    complete and detailed articles, it only contains the most interesting
    facts that you have learned from your research. It is a cross between
    information and personal interpretation, nevertheless being presented
    in a structured manner. That way, you can visit it later and remember
    those interesting fact that you learned during research, and maybe
    even share them with colleagues. It is not for note-taking, but it
    contains conclusions from cross-referenced sources and an unfolded
    understanding about a given topic.

    \begin{itemize}
    \item
      \textbf{\textit{\textless{}topic\textgreater{}.txt}}: Each topic
      will have its own page.
    \item
      \textbf{\textit{\textless{}topic\textgreater{}}:} A topic can also
      be extended into several subtopics, that are stored within the
      corresponding namespace.
    \end{itemize}
  \item
    \textbf{phd.txt:} The home page for your doctoral work, where you
    should list the title and statement of your thesis, along with initial
    planning or notes.
  \item
    \textbf{phd:} This is the main namespace, where everything about your
    doctoral work will be organized, including literature, datasets, and
    experiments.

    \begin{itemize}
    \item
      \textbf{bibliography.txt}: A categorized list of publication titles
      linking to their reading sheet.
    \item
      \textbf{bibliography:} This namespace should contain the pages for
      all reading sheets, named according to the title of the paper. A
      template for the reading sheet is provided for new pages within this
      namespace.

      \begin{itemize}
      \item
        \textbf{author.txt, year.txt, journal.txt, conference.txt,
        publisher.txt, institution.txt:} An alphabetically organized list
        of entities (authors, conferences, etc.).
      \item
        \textbf{author, year, journal, conference, publisher,
        institution:} Each entity associated with a publication will
        result in a page within these namespaces. The page for the entity
        will contain a list of backlinks, usually corresponding to paper
        titles, as well as optional information about the entity (e.g.,
        notes about an author). A template that will list backlinks is
        provided for each namespace.
      \item
        \textbf{list:} Sometimes, it is useful to build lists of papers to
        answer a research question (e.g., \emph{What PageRank variations
        and computation methods are there?}). This namespace serves this
        purpose, providing an organized location for such endeavors.
      \end{itemize}
    \item
      \textbf{collections.txt:} List all the internal links to datasets
      that you documented, categorized by type of data (e.g., networks)
      or, optionally, by another aspect (e.g., function, task).
    \item
      \textbf{collections:} This namespace will contain the individual
      pages for each dataset. The pages should be named according to the
      dataset name. A template is provided with basic collection metadata.
      This includes some statistics that might not apply to your dataset,
      but that can and should be easily replaced by other, more adequate
      descriptors.
    \item
      \textbf{experiments.txt:} List all the internal links to documented
      experiments, and organize information about the research
      methodology, or other top-level notes that might be required.
    \item
      \textbf{experiments:} This namespace will contain the individual
      pages for each experiment. The pages should be named according to
      the model or goal of the experiment. A template is provided with
      general metadata, as well as sections to document the challenges,
      dependencies, model versions, traces (or detailed steps), and
      research log entries, where you should take notes as the experiment
      is prepared and executed.
    \item
      \textbf{milestones.txt:} Particularly during the first year, you
      will be required to establish milestones for your doctoral work.
      This page lists these milestones (e.g., \emph{M1: Capture data from
      the brain in a non-invasive manner}), along with the top-level
      activities for each milestone (e.g., \emph{A: Obtain test data}) and
      the lower-level tasks for each activity (e.g., \emph{T1. Calibrate
      the machine}).
    \item
      \textbf{milestones:} This namespace will contain pages for each
      task, describing its goal and providing space for a research log
      that should use informal, or less formal, language.
    \item
      \textbf{resources.txt:} A list of hyperlinks to categorized web
      resources of interest to the thesis. This is similar to the
      \emph{bibliography.txt} page, although it is useful to separate them
      due to the difference in volume (i.e., we tend to collect a lot more
      links than papers, during research).
    \item
      \textbf{resources:} This namespace will contain optional pages to
      further document and analyze a given resource. Similarly to the
      reading sheets in the \emph{bibliography} namespace, we also provide
      a template for describing resources.
    \end{itemize}
  \item
    \textbf{\textit{\textless{}program\textgreater{}}.txt:} The main page
    for your master's or doctoral program, usually containing bureaucratic
    details, or general tasks.
  \item
    \textbf{\textit{\textless{}program\textgreater{}}:} The namespace
    where pages for each course should be included.
\end{itemize}

\subsection{Setup}
\label{setup}

Based on the practical experience we acquired while actively using and
testing this research management wiki, we have selected several useful
plugins that we list in the following table.

\begin{table}
  \renewcommand{\arraystretch}{1.3}

  \begin{adjustbox}{width=\linewidth}
    \begin{tabular}{lp{\linewidth}}
      \toprule
      Plugin & Description\\
      \midrule
      \href{https://www.dokuwiki.org/plugin:wrap}{Wrap} & Provides several overall useful features for spacing, coloring, or creating boxes.\\
      \href{https://www.dokuwiki.org/plugin:blockquote}{Blockquote} & Used to quote relevant passages from articles, usually while taking notes in the reading sheet.\\
      \href{https://www.dokuwiki.org/plugin:edittable}{EditTable} & Supports table editing with a spreadsheet-like user interface.\\
      \href{https://www.dokuwiki.org/plugin:tablewidth}{Table Width} & Enables the configuration of the width of individual columns in a table.\\
      \href{https://github.com/FyiurAmron/sortablejs}{sortablejs} & Adds the ability to sort by any column on a table.\\
      \href{http://www.dokuwiki.org/plugin:timestamp}{Timestamp} & Provides a consistently formatted timestamp with the current time (customizable).\\
      \href{http://www.dokuwiki.org/plugin:saveandedit}{saveandedit} & Adds a checkbox to the edit page, so that we can save and continue editing, without having to visit the rendered page first.\\
      \href{https://www.dokuwiki.org/plugin:NumberedHeadings}{Numbered Headings} & Adds the ability to automatically number headers.\\
      \href{http://www.dokuwiki.org/plugin:nspages}{nspages} & Used to list all pages within a given namespace (e.g., list all author or journal names).\\
      \href{https://www.dokuwiki.org/plugin:backlinks}{Backlinks} & Used to list pages that link to the current page (e.g., list all reading sheets for a given year or author).\\
      \href{https://www.dokuwiki.org/plugin:todo}{ToDo} & Provides a way to create checkable to-do items, as well as a way to list all to-do entries from a given namespace.\\
      \href{http://dokuwiki.org/plugin:folded}{folded} & Used to create a togglable block.\\
      \href{https://www.dokuwiki.org/plugin:mathjax}{MathJax} & Enables LaTeX math expressions.\\
      \href{https://www.dokuwiki.org/plugin:imagebox}{imagebox} & Provides a way to add captions to figures.\\
      \href{http://www.dokuwiki.org/plugin:move}{Move} & Adds an entry to the page's side menu, providing an easy way to move the page to a different namespace.\\
      \href{https://www.dokuwiki.org/plugin:dw2pdf}{dw2pdf} & It provides an easy way to download a PDF version of a page.\\
      \href{http://www.dokuwiki.org/plugin:refnotes}{RefNotes} & Enables the support for bibliographic references.\\
      \href{https://www.dokuwiki.org/plugin:backup}{BackupTool} & An easy way to download a backup of your wiki. You can also archive the `dokuwiki/` directory and restore it later, ensuring it the user and group are `www-data` (or, equivalently, UID and GID 33).\\
      \bottomrule
    \end{tabular}
  \end{adjustbox}
\end{table}

\subsection{Docker installation}
\label{docker-installation}

In order to facilitate adoption of this research methodology, we prepared a Docker-based distribution of a preconfigured instance of DokuWiki, that downloads and installs the previous extensions and skeleton (i.e., namespace structure, pages, and templates). This can be installed as follows:
\smallskip
\begin{lstlisting}[language=bash]
git clone https://github.com/jldevezas/research-wiki.git
cd research-wiki/
docker-compose up --build
\end{lstlisting}

\smallskip

\noindent You can then stop the server by pressing \textit{ctrl+c}.

\smallskip

\noindent Later on, to restart the instance, sending it to the background, just run:
\smallskip
\begin{lstlisting}[language=bash]
cd research-wiki/
docker-compose start
\end{lstlisting}

\smallskip

\noindent And you can stop it with:
\smallskip
\begin{lstlisting}[language=bash]
cd research-wiki/
docker-compose stop
\end{lstlisting}

\bigskip

In order to access your research wiki, visit \url{http://localhost:8080/} and login with username \textit{admin} and password \textit{admin} (don't forget to change it in \textit{Admin\ =\textgreater{}\ User\ Manager}). You can then start editing taking advantage of the multiple templates for the preconfigured namespaces within the proposed structure.

If instead you want to replicate this structure on a running DokuWiki instance, without relying on Docker, then you can simply install the suggested plugins, and extract the contents of the \textit{skeleton/} directory onto the \textit{data/pages/} directory in your DokuWiki instance.

The wiki instance managed by Docker is stored within the \textit{dokuwiki/} directory. According to the file structure of DokuWiki, pages and media files are stored within the \textit{dokuwiki/data/} directory. You might directly deploy the \textit{dokuwiki/} contents as a PHP website using an HTTP server of your choice (e.g., Apache HTTP Server\footnote{\url{https://httpd.apache.org/}}, nginx\footnote{\url{https://nginx.org/}}), and you can also store the \textit{dokuwiki/data/} directory as a backup.

\section{Automated meta-analysis}
\label{automated-meta-analysis}

We carried an automated analysis of the literature collected during the state of the art survey process. We prepared a Jupyter notebook based on the R kernel\footnote{\url{https://irkernel.github.io/}} to derive several statistics and plots from the entries added to the doctoral wiki about a subset of selected publications, following our systematic approach to documentation.

We first prepared a CSV file by scraping the doctoral wiki pages within the `phd:bibliography' namespace. The generated CSV file contains the following fields, which are empty when unavailable: `title'; `author' (multiple authors were separated by `\textbar{}'); `year'; `conference'; `core' (conference rank based on CORE 2018 data); `journal'; `scimago\_h\_index' (journal h-index as extracted from Scimago 2018 data); `institution'; `publisher'; and `review' (only available for some of publications in the doctoral wiki).

We used simple heuristics based on common expressions to obtain the conference name from the field corresponding to the proceedings entry. We then used the Jaccard index to match the conference name with an entry in CORE 2018, in an attempt to normalize it. Based on this approach, we were able to obtain the core ranking and the Scimago h-index for all conference and journal entries.

\subsection{Jupyter notebook}
\label{jupyter-notebook}

We provide a Jupyter notebook to help you analyze your research activity based on your wiki, mainly over the three dimensions we propose here: literature, collections and experiments. While outside the scope of this article, the notebook also supports the same analysis over a BibTeX file instead of the wiki.

In order to be able to run the analysis in the notebook, first you must install R\footnote{\url{https://www.r-project.org/}} and Jupyter\footnote{\url{https://jupyter.org/}}:
\smallskip
\begin{lstlisting}[language=bash]
  sudo apt-get install r-base
  pip install jupyter
\end{lstlisting}

\smallskip

\noindent Then, you must install the \textit{IRkernel} package in R. Just run \textit{R} from the command line and type:
\smallskip
\begin{lstlisting}[language=R]
  install.packages("IRKernel")
\end{lstlisting}

\smallskip

\noindent Finally, launch Jupyter notebook in the notebook directory that is provided with the \textit{research-wiki} GitHub repository\footnote{\url{https://github.com/jldevezas/research-wiki/}}:

\begin{lstlisting}[language=bash]
  cd research-wiki/notebook/
  jupyter notebook
\end{lstlisting}

\begin{figure}
  \centering
  \hspace{4.5em}\includegraphics[width=.725\linewidth]{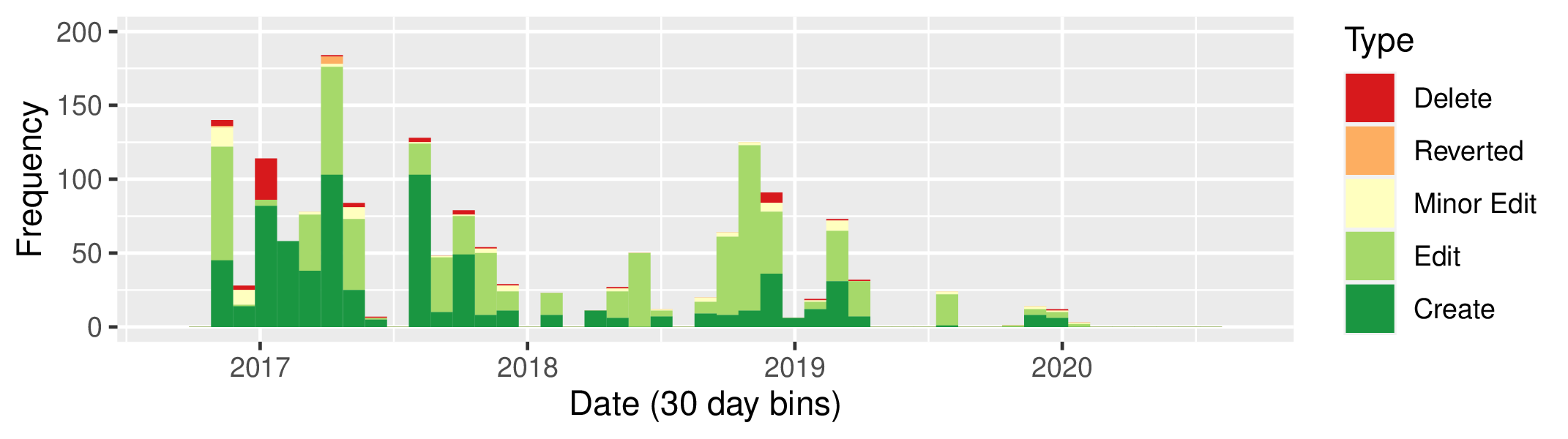}
  \caption{Wiki changes over time for the literature review pages.}
  \label{fig5}
\end{figure}

\begin{figure}
  \centering
  \includegraphics[width=.625\linewidth]{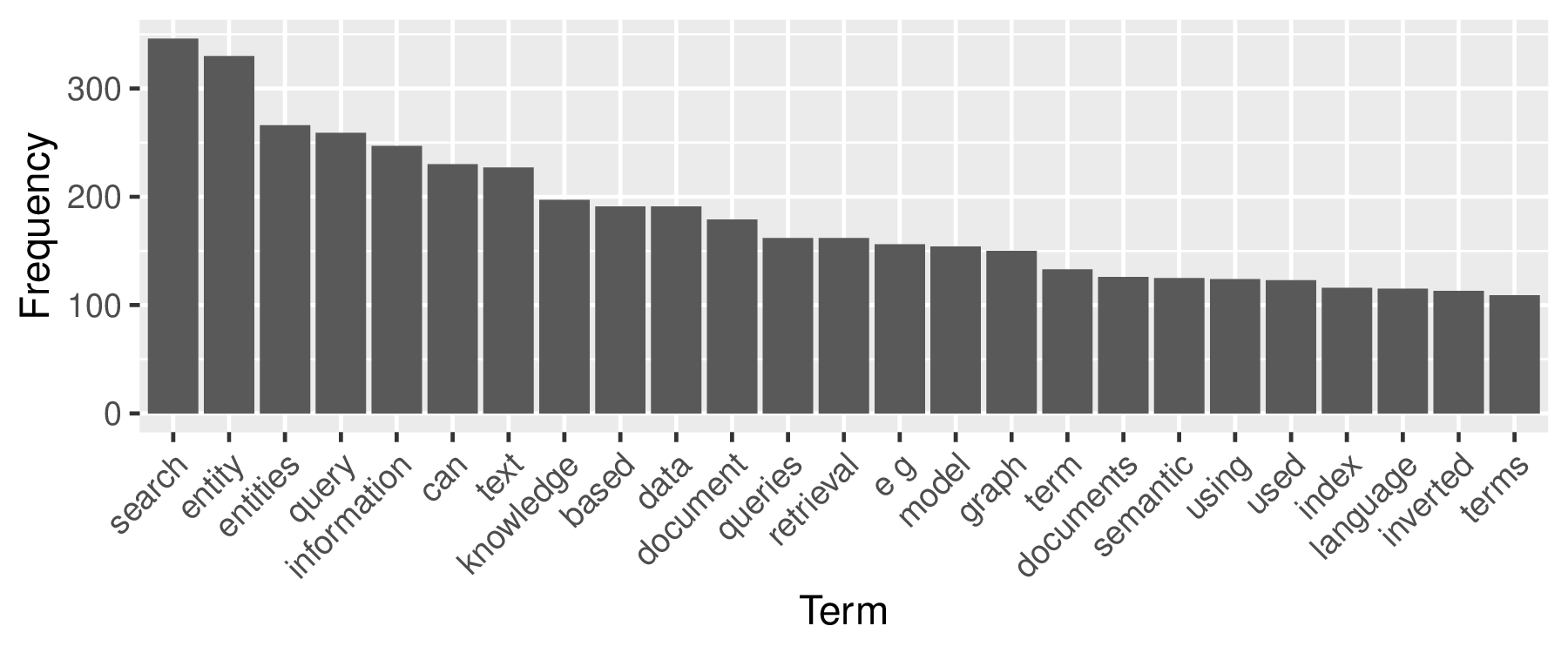}
  \caption{Term frequency distribution for the vocabulary used in the
  reading sheets during literature survey.}
  \label{fig6}
\end{figure}

Follow the instructions in the notebook to run your analysis. All plots will be saved to the \textit{output/} directory. Among others, there you will find a plot for the wiki changes to pages within the \textit{phd:bibliography} namespace, as illustrated in Figure~\ref{fig5}, as well as within the \textit{phd:collections} and \textit{phd:experiments} namespaces. You will also find statistics about authors, journals, conferences, publication years, or even term frequency distributions for your own notes, taken during the literature review stage, as shown in Figure~\ref{fig6}.

\section{Closing notes}
\label{closing-notes}

At the end of four years of work, we can safely say that relying on a wiki to systematically document our research work was an overall good decision. One thing we would have done differently would be the organization of experiments. While, in theory, it makes sense to create individual pages per experiment, in practice experiments carry on and grow vertically (e.g., extensions to a model), more so than horizontally (e.g., new models). Accordingly, perhaps a structure relying on sub-experiments would make sense. In practice we created separate log entries for each idea that spawned from the main experiment, some of which materialized into new experiments that shared most of the information about the main experiment. The best or most general approach to deal with this is still unclear, however the simple knowledge that experiments will grow vertically will enable researchers to better organize their experiment sheets.

While it is important to ensure a solid organization scheme, it is also important to safeguard the doctoral wiki as a personal space, where you can draft your ideas and document the process. The output is not always presentable for an outside audience and, sometimes, not even your team, but freedom to write and rewrite and drop ideas is fundamental for research. The wiki is your notebook and it should be allowed to be messy. Work should be organized elsewhere, in papers or reports, but the wiki should provide freedom to develop and support the organization of ideas.

\begin{acks}
  Thank you to Professor João Correia Lopes, who is also an enthusiastic wiki user, for sharing his favorite DokuWiki plugins, many of which we adopted for the research management wiki we present here.

  \medskip

  \noindent José Devezas was supported by research grant PD/BD/128160/2016, provided by the Portuguese national funding agency for science, research and technology, Fundação para a Ciência e a Tecnologia (FCT), within the scope of Operational Program Human Capital (POCH), supported by the European Social Fund and by national funds from MCTES.
\end{acks}

\bibliographystyle{ACM-Reference-Format}
\bibliography{research_wiki.bib}

\end{document}